\documentclass[a4paper,fleqn,usenatbib]{mnras}

\usepackage{mathptmx}

\usepackage[T1]{fontenc}
\usepackage{ae,aecompl}

\usepackage{graphicx}	
\usepackage{amsmath}	
\usepackage{amssymb}	
\usepackage{BibDef}
\usepackage{geometry}
\usepackage{lipsum}

\newcommand{\msun}{\ensuremath{\, \mathrm M_{\sun{}}}}

\newcommand{\ihat}{\hat{\textbf{\i}}}
\newcommand{\jhat}{\hat{\textbf{\j}}}
\newcommand{\khat}{\hat{ \bf k}}
\newcommand{\vect}[1]{{\mathbf{ #1}}}

\title[SN kicks in the GC]%
{Supernova Kicks and Dynamics of Compact Remnants in the Galactic Centre}

\author[Bortolas et al.]{
Elisa Bortolas,$^{1,2}$\thanks{E-mail: elisa.bortolas@oapd.inaf.it}
Michela Mapelli,$^{1,3}$, Mario Spera$^{1,3}$
\\
$^{1}$INAF, Osservatorio Astronomico di Padova, Vicolo dell'Osservatorio 5, I-35122, Padova, Italy\\
$^{2}$Dipartimento di Fisica e Astronomia ''Galileo Galilei'', Universit\`a di Padova, Vicolo dell'Osservatorio 3, 35122 Padova, Italy\\
$^{3}$INFN, Sezione di Milano-Bicocca, Piazza della Scienza 3, 20126 Milano, Italy\\
}
\date{Accepted XXX. Received YYY; in original from ZZZ}

\pubyear{2016}

\begin{document}
\label{firstpage}
\pagerange{\pageref{firstpage}--\pageref{lastpage}}
\maketitle

\begin{abstract}
The Galactic Centre (GC) is a unique place to study the extreme dynamical processes occurring near a super-massive black hole (SMBH). Here we investigate the role of supernova (SN) explosions occurring in massive binary systems lying in { a disc-like structure within} the innermost parsec. We use {a regularized  algorithm }  to simulate { $3\times{}10^4$ isolated three-body systems composed of a stellar binary orbiting the SMBH. We start the integration when} the primary member undergoes a SN explosion, and we analyze the impact of SN kicks on the orbits of stars and compact remnants. 
We find that SN explosions scatter { the lighter stars  in the pair }
on completely different orbits, with higher eccentricity and inclination. In contrast, stellar-mass black holes (BHs) and massive stars retain memory of the orbit of their progenitor star.  Our results suggest that SN kicks are not sufficient to eject BHs from the GC. 
{ We thus predict that all BHs that form {\it in situ} in the central parsec of our Galaxy remain in the GC, building up a cluster of dark remnants}. In addition, the change of NS orbits induced by SNe may partially account for the observed dearth of NSs in the GC.
About 40 per cent of remnants stay bound to the stellar companion after the kick; we expect up to 70 per cent of them might become X-ray binaries through Roche-lobe filling. Finally, the eccentricity of some light stars becomes $>0.7$ as an effect of the SN kick, producing orbits similar to those of the G1 and G2 dusty objects.
\end{abstract}



\section{Introduction}\label{sec:intro}
The Galactic Centre (GC) is an overwhelmingly crowded environment: it hosts a large population of late-type stars, few hundreds of early-type stars, several gas structures, and a $\sim{}4\times 10^6 \msun$ super-massive black hole (SMBH), associated with the radio source SgrA$^\ast$ (see, e.g. \citealt{MapelliGualandris2016} for a recent review). 

Being the closest galactic nucleus to us, the GC has been one of the most studied places in the Universe over the last decades, but still  poses a plethora of questions.  
 First of all, star formation is not expected to take place `in normal conditions' within the central few parsecs of the Milky Way, because molecular clouds are disrupted by the 
extreme gravitational field of the SMBH \citep{Sanders1998}. 
Nevertheless, a population of $\sim{}30$ young ($\sim{}20-100$ Myr) B-type stars (the S-stars) are found within 0.04 pc 
from SgrA$^\ast$ \citep{Morris1993,Ghez2003}. Moreover, several hundred  young ($\sim{}2-6$ Myr, \citealt{Lu2013}) early-type stars lie in the innermost parsec. About $20\%$ of them belong to the so-called clockwise (CW)  disc, a thin-disc structure extended between $\sim0.04$ and 0.13 pc from SgrA$^\ast$. CW disc members are found to follow a top-heavy (TH) mass function, best fitted by a power law with index $\alpha{}\sim{}1.7$ \citep{Lu2013}. 

Many scenarios have been proposed to explain the formation of the early-type stars, and this issue is still debated. According to the migration scenarios, the young stars formed out of the central parsec, and then sank to the GC by dynamical mechanisms { (e.g. \citealt{Kim2003,Kim2004,Fujii2008,Fujii2010,Perets2009,Perets2010}, but see \citealt{Petts2017} for the most recent discussion of the issues connected with this scenario)}, while in the in situ scenarios
star formation can be triggered by the disruption of a giant molecular cloud (e.g. \citealt{Sanders1998,Bonnell2008,Hobbs2009,Alig2011,Mapelli2012,Alig2013,Mapelli2013}).

Moreover, two faint dusty objects, named G1 and G2, have been observed in the vicinity of the SMBH, on highly eccentric orbits \citep{Clenet2004a, Clenet2004b, Clenet2005, Ghez2005, Gillessen2012,Gillessen2013,Pfuhl2015}. In particular, G2 approached the SMBH to $\sim{}200$ AU  in Spring 2014, during periapsis passage, and was not completely disrupted by the SMBH \citep{Witzel2014}. G1 and G2 have been extensively studied since their first discovery: they have been proposed to be gas clouds (e.g. \citealt{Schartmann2012}), dust-enshrouded low-mass stars (e.g. \citealt{Burkert2012,Ballone2013,Scoville2013,Prodan2015}), tidally-disrupted stars (e.g. \citealt{Guillochon2014}), or planetary embryos and/or proto-planetary discs (e.g. \citealt{Murray2012,Mapelli2015,Trani2016}), but their origin remains unclear.  

Another major issue about the GC is the observed cored profile of late-type ($>1$ Gyr) red giant stars  \citep{Buchholz2009,Do2009,Do2013,Aharon2015}. If the old stellar population in the GC were relaxed, we would expect it to follow a cusp profile { with slope $\sim{}-7/4$ } \citep{Bahcall1976}, but red giant stars exhibit a cored  profile, thus hinting to a possible \emph{missing cusp problem}. { A genuine cored distribution might be the result of the infall of star clusters} \citep{Antonini2012,Perets2014}, or of a recent SMBH binary merger \citep{Merritt2010}. Alternatively, red giant stars might not be representative of the late-type stellar distribution, { as suggested in  a series of recent studies considering the population of faint giants and sub-giants \citep{Gallego-Cano2017,Schodel2017,Baumgardt2017}: if only these populations are considered, the stellar cusp seem to be present, even if it appears to be shallower than expected}.  
The { still apparent lack of bright } red giants might be explained as the outcome of stellar collisions, removing mainly red giants \citep{Genzel1996,Alexander1999, Bailey1999, Dale2009}. 

Pulsars are also missing in the GC. Based on the large number of massive stars, we expect a large population of neutron stars (NSs) in the central parsec (up to few thousands in the innermost $\sim 10^{-2}$ pc, according to \citealt{Pfahl2004}), but only one was observed:  PSR J1745-2900, a magnetar orbiting SgrA$^\ast$ with $\lesssim{}0.1$ pc of projected separation (\citealt{Kennea2013,Mori2013,Rea2013}, but see also \citealt{Wang2006,Muno2008,Ponti2016} for other possible candidates). The observation of the magnetar implies that the lack of other pulsar detections is an intrinsic dearth of pulsars in the central parsec, rather than an effect of  strong interstellar scattering screen toward the GC. Moreover, the magnetar formation efficiency might be higher in the GC than in other places \citep{Dexter2014}.




In this paper, we investigate the missing cusp problem and the missing pulsar problem by means of a simple argument: supernovae (SNe) occurring in binary systems in the central parsec might reshuffle the orbits of pulsars, other dark remnants, and old red giant stars. Moreover, SN kicks in binary systems might push low-mass stars into highly eccentric orbits, such as the ones of G1 and G2.  SN kicks might be very large \citep{Hobbs2005}, and strongly affect the orbit of a binary star, in many cases breaking up the binary. 

Thus, we use { three-body simulations  } to study the dynamical effects of SNe occurring in binary stellar systems within a disc structure similar to the CW disc.  We show that SN kicks reshuffle the orbit of the binary and contribute to sculpt the density profile of the GC. 

\section{Methods}\label{sec:methods}
To study the evolution of GC binaries after SN explosions, we run 30,000 three-body encounters of systems composed of a SMBH and a stellar binary. The mass of the SMBH is fixed to $4.3\times{}10^6\,{}M_\odot$ \citep{Gillessen2009}.


The binary centre of mass is initially assumed to sweep a Keplerian orbit around the SMBH, with eccentricity and semi-major axis in agreement with recent observations of the CW disc: the eccentricity $e$ is drawn from a Gaussian distribution centred in $\langle{}e\rangle{}=0.3\pm{}0.1$ \citep{Yelda2014},  while the semi-major axes $a$ follow the distribution $f(a)\propto  a^{0.07}$ \citep{Do2013}, and we chose the mean anomaly of the orbit from a uniform distribution between 0 and $2\pi$; this ensures that the surface density profile of stars is $\Sigma(r)\propto r^{-0.93}$, consistent with the surface density distribution of the early type stars in the disc \citep{Do2013}; {  $a$  can vary in the range [$0.04, 0.13$] pc or [$0.001, 0.13$] pc depending on the simulation set, as detailed below.}

The eccentricity $e_\ast$ of the stellar binary is randomly extracted from the thermal eccentricity distribution $f(e_\ast{}){\rm d} e_\ast{}=2e_\ast{}{\rm d}e_\ast{}$ \citep{Jeans1919}, while the semi-major axis $a_\ast{}$ distribution takes the form $f(a_\ast{}) {\rm d} a_\ast{}\propto a_\ast{}^{-1}{\rm d}a_\ast{}$, i.e. follows the so-called \emph{\"Opik's law}  \citep{Opik1924}.  The semi-major axis of the binary system spans from a minimum of 0.2 AU to a maximum distance equal to $0.4\,{}r_{\rm j}$, where $r_{\rm j}=a\,{}(1-e)\,{}(\tfrac 1 3 M_{b,i}/M_{\bullet})^{1/3}$ is the Jacobi radius (where $M_{b,i}=M_{1,i}+M_{2}$, $M_{1,i}$ is the mass of the primary before the SN explosion, $M_2$ is the mass of the secondary, and $M_\bullet$ is the mass of the SMBH)\footnote{ The resulting initial orbital period of the stellar binaries ranges from $\sim 0.1$ yr to $\sim 10^3$ yr, and about one fourth of them has a period $<1$ yr.}. This ensures by construction that 
the binary star will not break by tidal forces at the closest passage with the SMBH, as shown in \citet{Trani2016b}. We also ran a sub-sample of 3000 three-body simulations without SN kicks, to check that the binary does not break if SN kicks are not included in the simulations. We found that $a_\ast$ and $e_\ast$ change by less than $10^{-5}$ in the test simulations without SN kicks, indicating that our initial conditions are stable against tidal forces.

Since a large fraction of stars (especially massive stars) form in binary systems (e.g. \citealt{Sana2011b}), many CW disc members are expected to be binary stars. \citet{Gautam2016} found that the binary fraction of early-type main sequence stars within the CW disc is at least 32\% (within the 90\% confidence level). Stars with mass in excess of $\sim 9\msun$ are thought to undergo SN  explosions after about $3 - 50$ Myr, depending on their initial masses. Given the young age of stars in the CW disc ($\sim{}2.5 - 6$ Myr, \citealt{Lu2013}), most SN events will occur in the next Myrs. However, we use the orbital properties of the CW disc as initial conditions, because  it is reasonable to expect that a series of CW-disc like star-forming events have occurred in the past and may occur in the future (it is unlikely that we live in a special time for the history of the GC, \citealt{Loose1982}). 

\begin{figure}
\includegraphics[trim={0.5cm 0cm 7.5cm 20cm},width=\columnwidth]{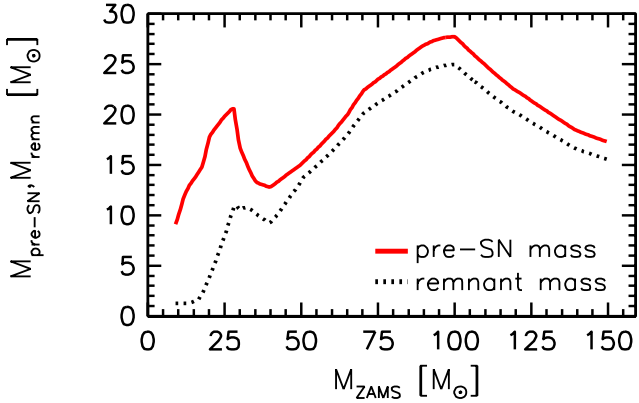} 
    \caption{Red solid line: pre-SN mass of a star as a function of its ZAMS mass according to the PARSEC stellar evolutionary tracks { \citep{Bressan2012,chen2015}}. Black dotted line: mass of the SN compact remnant as a function of the the progenitor ZAMS mass according to \citet{Spera2015}; we assume the  delayed SN model.}
    \label{fig:mfunc}
\end{figure}

The masses of the two stars are computed as follows: the zero age main sequence (ZAMS) mass of the more massive (primary) binary component ($M_{1,\rm ZAMS}$) is randomly distributed in the range $[9,150]\msun$, { sampling a TH mass function} (${\rm d}N/{\rm d } m \propto m^\alpha$, with $\alpha=-1.7$), in agreement with the observations of the GC \citep{Lu2013}. The ZAMS mass of the less massive (secondary) binary component  ($M_{2,\rm ZAMS}$) is either extracted from the TH mass function in the range $[0.1,\,{}150]\msun$ (simulation sets A and C) or assumed to be equal to 0.9 times the ZAMS mass of the primary star (simulation set B), as  detailed below.  We then evolve each primary star from $M_{1,\rm ZAMS}$ to its pre-SN mass $M_{1,i}$, according to the PARSEC stellar evolutionary tracks { \citep{Bressan2012,chen2015}};  the pre-SN mass of a star as a function of its ZAMS mass is shown in Figure~\ref{fig:mfunc}. Similarly, we calculate the mass of the secondary star at the time of the SN explosion of the primary ($M_2$) using the PARSEC stellar evolutionary tracks for a star of ZAMS mass $M_{2,\rm ZAMS}$. 


We start the simulations at the time of the SN explosion, by giving the binary a SN kick.  The mass of the compact remnant is computed using the  delayed SN model \citep{Fryer2012}, { according to \citet{Spera2015}}. We distinguish between NSs and  stellar-size black holes (BHs) depending on whether their mass is respectively below or above $3\msun$. The mass of the SN relic as a function of the progenitor ZAMS mass is shown in Figure \ref{fig:mfunc}.  
The distribution of SN kicks is uncertain, especially for BHs (see e.g. \citealt{Gualandris2005,Fragos2009,Repetto2012,Janka2013,Beniamini2016,Mandel2016}). For NSs, we adopt  the kick distribution reported by \citet{Hobbs2005}, i.e.  a Maxwellian distribution with one-dimensional variance equal to $\sigma=265$ kms$^{-1}$, derived on the basis of peculiar motions of 233 pulsars. If the remnant is a BH, the kick velocity is then normalized to the mass of the remnant, assuming linear momentum conservation. The natal kick is given to the binary members as detailed in Appendix \ref{app}.

\begin{table}
  \centering
  \caption{Initial conditions and differences between the three sets of simulations. }
  \label{tab:sets}
  \begin{tabular}{lccc} 
    \hline
                   &  set A        & set B         & set C \\
    \hline
    $M_{2,\rm ZAMS}$   & TH, $0.1-150\msun$ & $0.9\times M_{1,\rm ZAMS}$ & TH, $0.1-150\msun$\\
    CW disc & $0.04-0.13$ pc & $0.04-0.13$ pc &$0.001-0.13$ pc \\
    $N$ & 10,000 & 10,000 & 10,000 \\
    $\Delta t_{\rm sim}$ & 1 Myr & 1 Myr & 1 Myr \\   
    \hline
  \end{tabular}
\begin{flushleft}
{\footnotesize Each row indicates, for the three sets, (i) how the ZAMS mass of the secondary component ($M_{2,\rm ZAMS}$) is chosen, (ii) the extension of the CW disc, (iii) the number $N$ of simulations in each set, (iv) the  time $\Delta t_{\rm sim}$ over which each system was evolved. }
\end{flushleft}
\end{table}

We performed three sets of three-body simulations composed of 10,000 runs
 each:  (i) in set~A, the ZAMS mass of the secondary star is extracted from 
 the TH mass function between 0.1 and $150\msun$ and the disc inner and
  outer limits are 0.04 and 0.13 pc, respectively; (ii) set~B differs from 
  set~A as the ZAMS mass of the secondary is set equal to 0.9 times the 
  ZAMS mass of the primary; we explored this possibility due to the 
  uncertainty in the mass correlation of binary stars 
  \citep{Kobulnicky2007,Sana2011}; (iii) in set~C the ZAMS mass of the 
  secondary is computed as in set~A, but the disc is assumed to span from 
  0.001 to 0.13 pc, according to \citet{Chen2014}, who suggested that the 
  CW disc inner edge was closer to SgrA$^\ast{}$ in the past. The 
  differences among the three sets and the parameters of the simulations 
  are summarized in Table \ref{tab:sets}. {  We chose to perform 10$^4$ 
  runs per each set of simulations because each set aims at reproducing the 
  population of the CW disc, which has been estimated to be as massive as a 
  few$\times 10^4 \msun$ \citep{Lu2013}.  Moreover, we checked that 10$^4$ 
  runs per each simulation set are sufficient to filter out statistical 
  fluctuations.}

Each three-body  system is evolved from the moment of the SN explosion and for $\Delta t_{\rm sim}$=1 Myr; this timescale ensures that stars still bound to the SMBH generally complete $\sim{}100-1,000$ orbits by the end of the simulation, thus they totally absorb the effect of the kick. On the other hand, our simulations cannot account for the two-body relaxation, which can affect stellar orbits of a CW disc structure within $10-100$ Myr \citep{Subr2014}; thus a choice of $\Delta t_{\rm sim}> 1$ Myr would not be beneficial.

{ The simulations are evolved } by means of a fully regularized N-body code that implements the Mikkola's algorithmic regularization (MAR, \citealt{Mikkola1999a,Mikkola1999b}). This code is designed for studying the dynamical evolution of few-body systems in which strong gravitational encounters are frequent and the mass ratio between the interacting objects is large. The MAR scheme removes the singularity of the two-body gravitational potential for $r\rightarrow{}0$, by means of a transformation of the time coordinate (see \citealt{Mikkola1999a} for details). Our implementation uses a leapfrog scheme in combination with the Bulirsh-Stoer extrapolation algorithm. The code integrates the equations of motion employing relative coordinates by means of the so called chain structure. This change of coordinates reduces round-off errors significantly \citep{Aarseth2003}: { the relative error in the energy conservation of our simulations is always below $10^{-10}$}. The code can be used as a stand-alone regularized N-body code or as a module of the direct N-body code HiGPUs-R (Spera, in preparation; see \citet{Capuzzo2013} for the current  version of HiGPUs). 

\begin{figure}
\includegraphics[trim={0cm 0cm 9.5cm 16.3cm},width=\columnwidth]{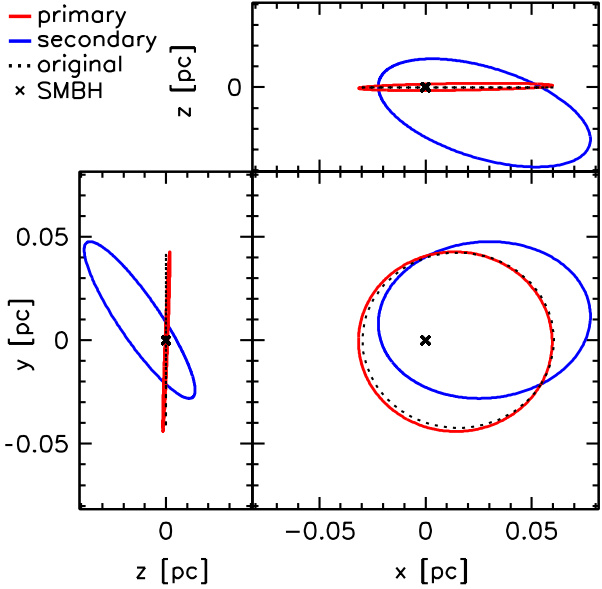} 
    \caption{ Orbit of the binary components in a selected simulation of set C. Black dotted line: pre-SN orbit of the centre of mass of the binary around the SMBH; red solid line: orbit of the compact remnant (a NS with mass $=1.3$ $\msun$) around the SMBH, after the SN explosion (which breaks the binary); blue solid line: orbit of the secondary star (mass $=0.1$ $\msun$) around the SMBH, after the SN explosion; black cross: position of the SMBH. Initial orbital parameters of the binary centre of mass: semimajor axis $a\approx45.0$ mpc, eccentricity $e\approx0.34$; orbital parameters of the secondary binary member at the end of the integration: semimajor axis $a\approx 53.0$ mpc, eccentricity $e\approx0.60$, difference in the orbital inclination $\approx 32$ degrees; the NS changes its orbital parameters by $\lesssim 10\%$ by the end of the run, as its semimajor axis becomes $\approx$45.6 mpc, its eccentricity $\approx 0.38$ and its inclination $\approx 2$ degrees.}
    \label{fig:orbit}
\end{figure}

{ Figure~\ref{fig:orbit}  shows an example of the binary orbit before and after the SN kick in set C: two stars of 9 and 0.1 $\msun$ orbit the SMBH with semimajor axis  $a \approx 45.0$ mpc and eccentricity $e \approx 0.34$. In this particular case the binary breaks, but both  members remain bound to the SMBH. The secondary star settles on a different orbit with respect to the initial one (with semimajor axis $\approx 53.0$ mpc, eccentricity $\approx 0.60$ and change in the orbital inclination of $\approx 32$ degrees) while the 1.3 $\msun$ NS produced by the SN only experiences a small ($\lesssim 10\% $) change in its orbital parameters. }

\section{Results}\label{sec:res}

\begin{figure*}
\includegraphics[trim={1.5cm 0cm 7.5cm 16cm},width=.3\textwidth]{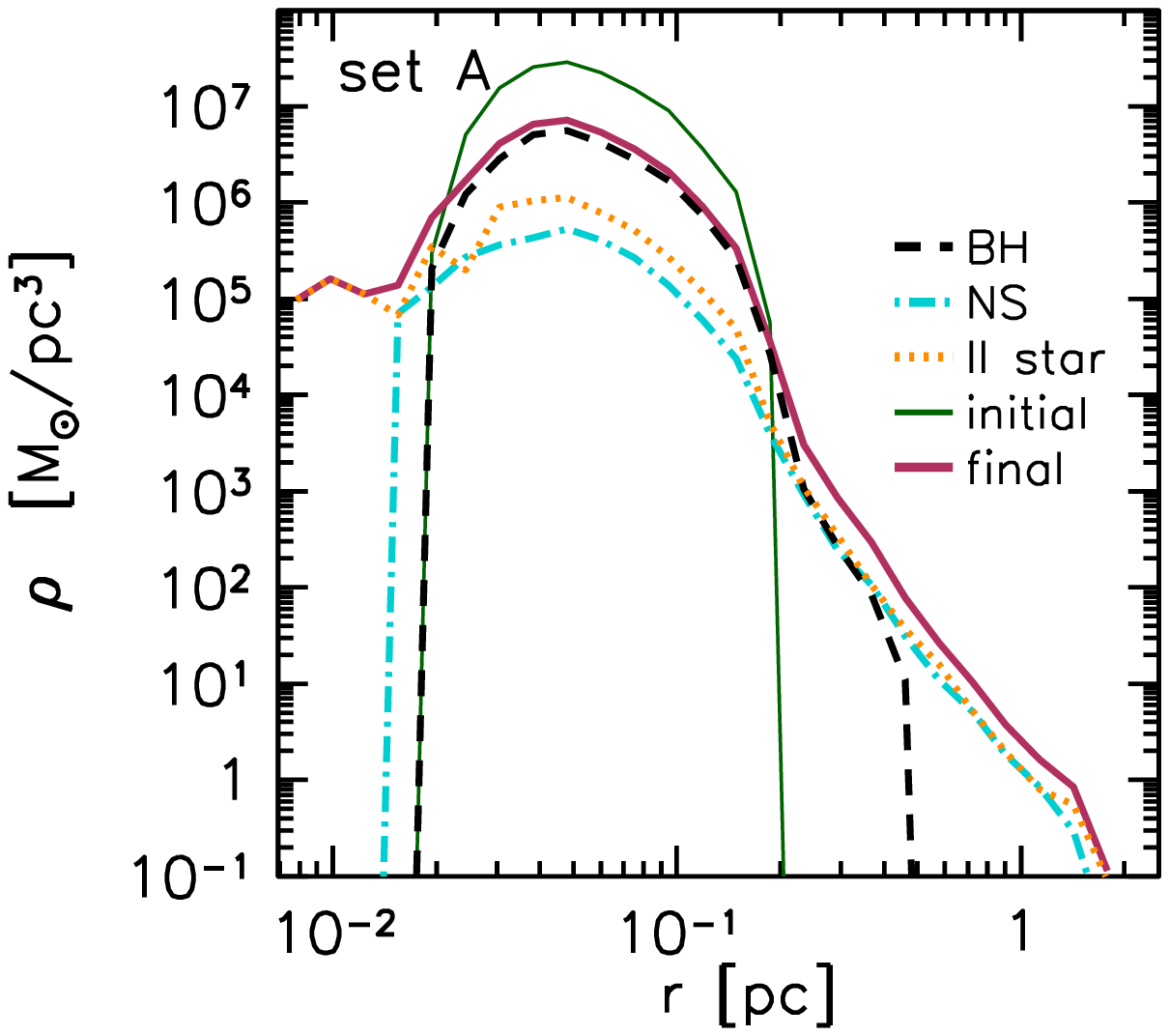} 
\includegraphics[trim={1.5cm 0cm 7.5cm 16cm},width=.3\textwidth]{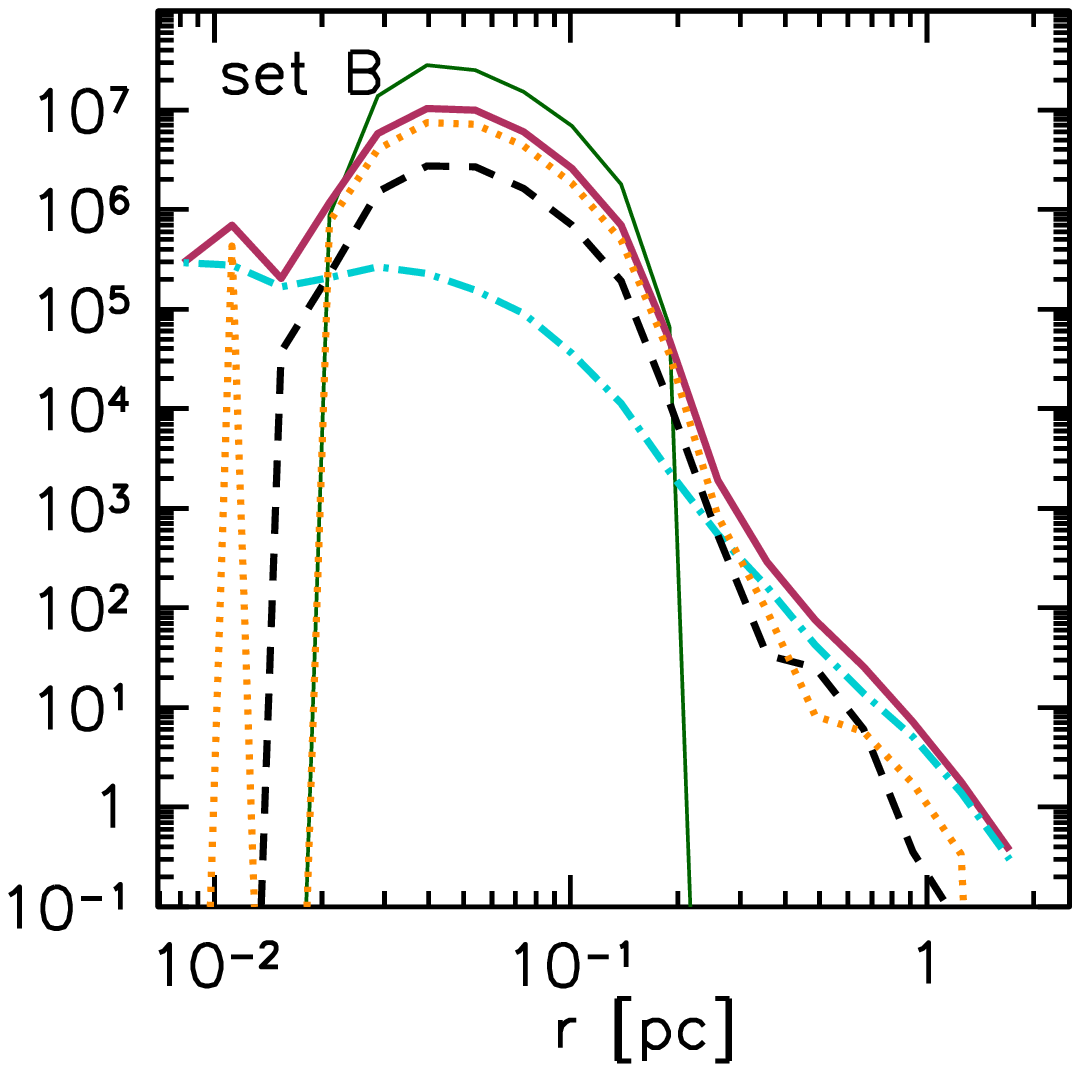} 
\includegraphics[trim={1.5cm 0cm 7.5cm 16cm},width=.3\textwidth]{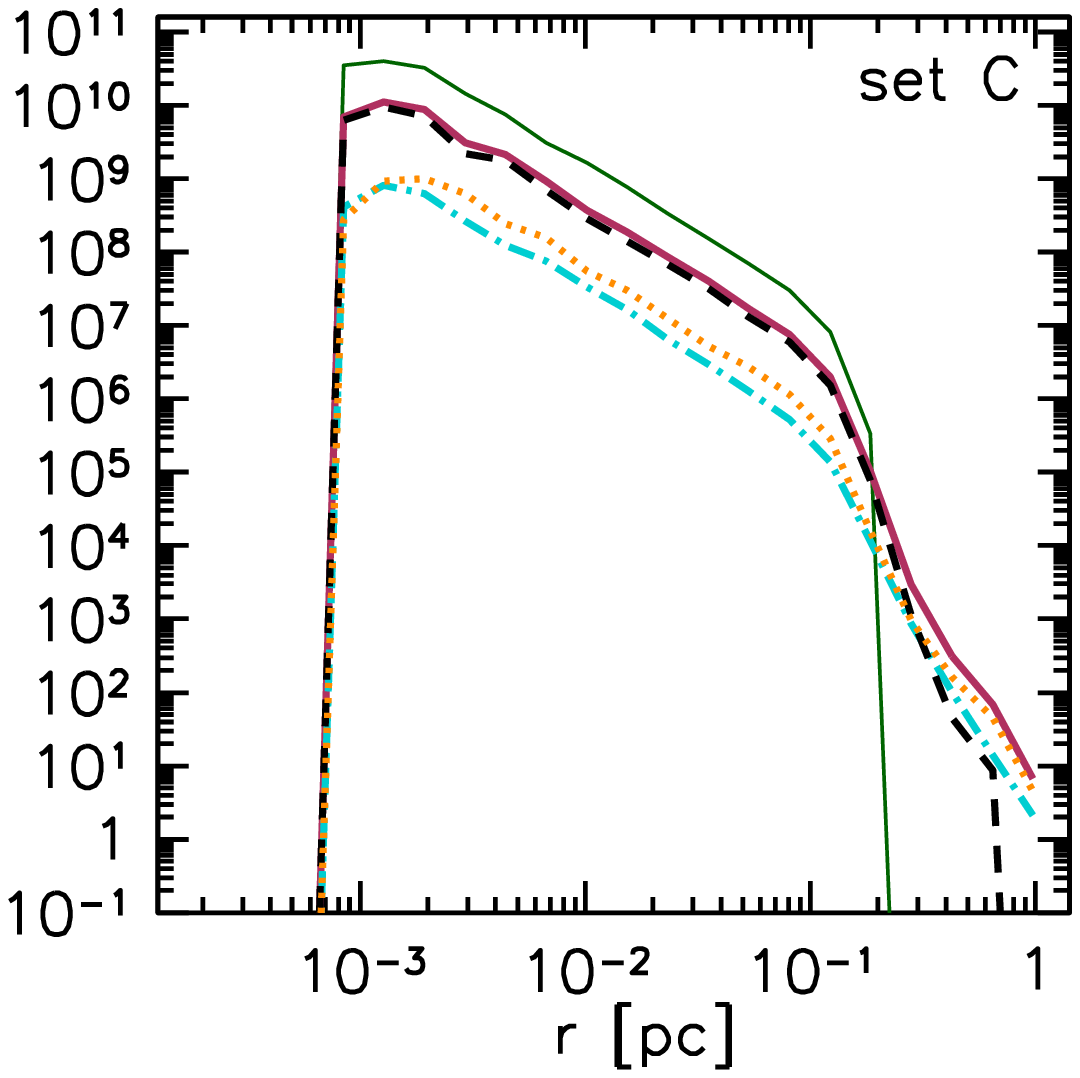} 
    \caption{Mass density profile of stars and compact remnants { reconstructed from the individual three-body runs}, for set~A (left), B(centre), C(right). Dark green solid thin line: total initial distribution; dark red solid line: total final distribution; black dashed line: BHs at 1 Myr; light blue dash-dotted line: NS at 1 Myr; orange dotted line: secondary stars at 1 Myr. The density profiles are normalized assuming an initial total mass equal to $10^5\msun$. }
    \label{fig:dens}
\end{figure*}

In this Section, we present the results of our three-body simulations. { Figure~\ref{fig:dens} shows the density profile of the simulated binaries reconstructed from the individual three-body experiments, by considering the distance of each star from the SMBH at time $t=1$ Myr (i.e. the end of the simulations).} Figure~\ref{fig:hist} shows the orbital parameter distribution of objects that are still bound to the SMBH at the end of the simulation. 
The semi-major axes, eccentricities and pericentre radii are computed with respect to the SMBH, while the inclination is computed with respect to the CW disc. 
In the plots, we distinguish between BHs ($\sim{}54\%$ of the remnants), NSs ($\sim{}46\%$ of the remnants) and secondary stars. We stress that the large number of BHs is due to the assumed TH mass function. The distribution of BHs resembles the pre-SN distribution, implying that most BHs are anchored to their initial orbit and SN kicks are not efficient in scattering them in different regions of the phase space. In contrast, NSs are more easily scattered by the SN explosion and spread in the innermost and outermost regions of the GC. This is valid for all the sets of simulations.

 In both sets~A and C,  $75-80\%$ of the mass within $0.1-1$ pc  is in the form of BHs. From Figure~\ref{fig:dens} it is also clear that in set~A and C BHs dominate the final mass distribution for a large range of radii, i.e. within $\sim 0.02-0.2$ pc in set~A and $\sim 5\times 10^{-4}-0.2$ in set C. { We stress that our simulations were run only for 1 Myr and that they only account for three bodies at a time: we cannot say anything about the effect of two-body relaxation on the BHs confined in the central pc. { However, it is likely that the BHs will relax their distribution near the SMBH within $\sim 100$ Myr; the relaxation time within the disc is computed as described in \citet{Subr2014}%
\footnote{
As a matter of fact, the relaxation time-scale of the old and almost isotropic stellar component within the SMBH influence radius has been estimated to be $\sim10$ Gyr or even longer \citep[][fig. 3]{Merritt2010}. However our paper focuses on the young population of stars in the CW disc: due to their coherent motion, stars in a disc self-relax on a much shorter time-scale $T_{\langle{}e^2\rangle{}}$, that can be computed as the time over which the mean-square-eccentricities of stars grow from zero to a given $\langle{}e^2\rangle{}$ \citep{Stewart2000}. If we assume a CW-disc like structure and we use ${\langle{}e^2\rangle{}}^{1/2}\approx0.3$,   it follows that $T_{\langle{}e^2\rangle{}}\sim 100$ Myr \citep[][equation 4]{Subr2014}.
%
}%
.}

The behaviour of the secondary stars strongly depends on their masses: in set~A and C, the secondary binary member is a low-mass star in most cases ($\sim 90\%$ secondary stars have masses lower than $3\msun$) and is easily scattered on a completely new orbit after the SN kick, while in set B the secondary is a high-mass star ($\gtrsim 9\msun$) and keeps memory of its initial orbit. In set~B, secondary stars are much more massive than the remnants of primary stars, thus they dominate the mass distribution within $\sim 0.02-0.5$ pc. This would be true only within a short timescale, as most of the companion stars in set~B will undergo a SN explosion within a few Myr and they will turn into compact remnants.

\begin{figure*}
\includegraphics[trim={0cm 2cm 0cm 10cm},width=\textwidth]{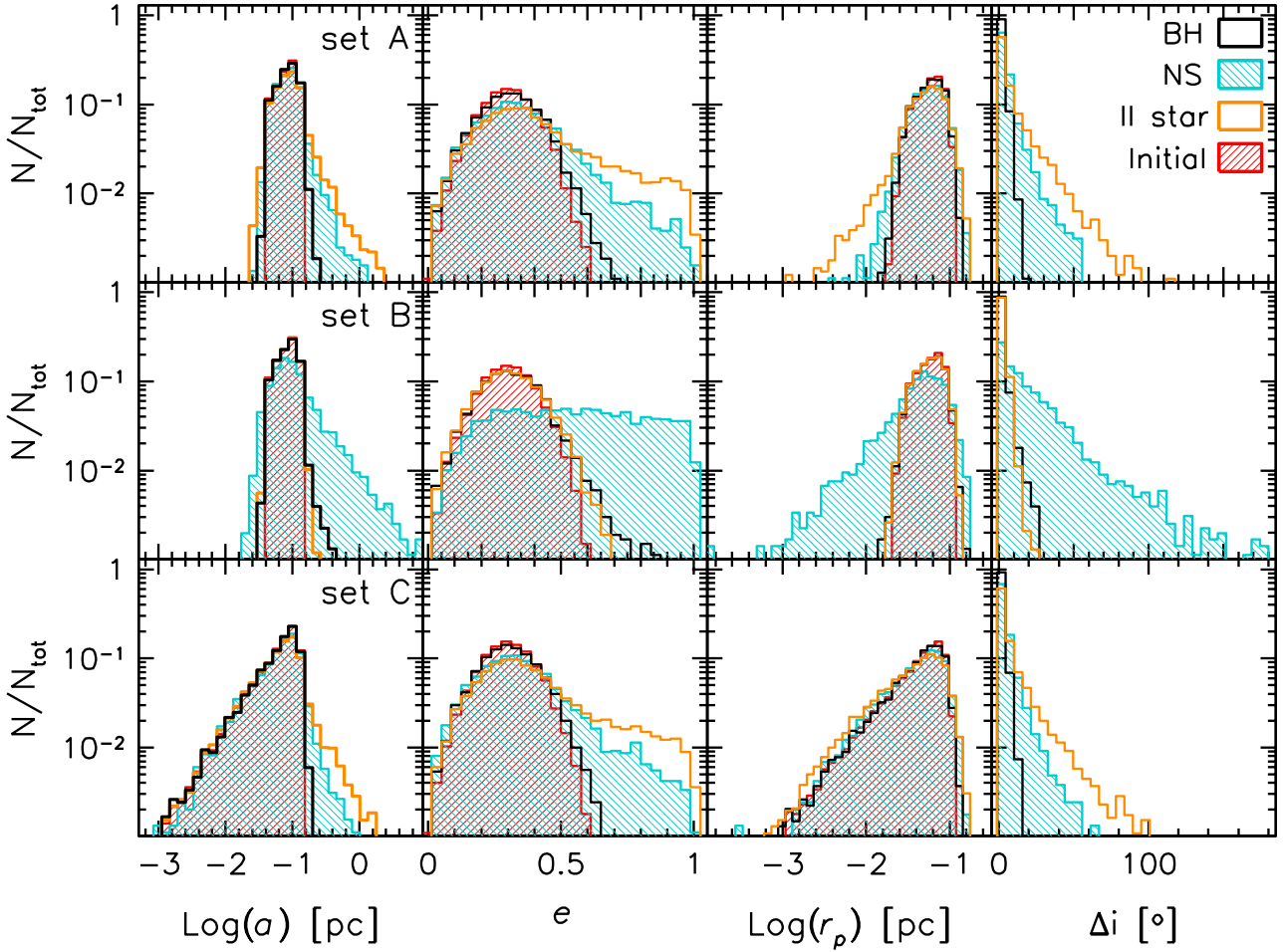} 
    \caption{From left to right: distribution of semi-major axes, eccentricities, pericentre radii, and inclinations for objects still bound to the SMBH. Inclinations are calculated with respect to the initial orbit of the binary (i.e. the CW disc plane). Top row: set~A; central row: set~B; bottom row: set~C. Red shaded histograms: initial distribution; black histograms: BHs at 1 Myr; light blue dashed histograms: NSs at 1 Myr; orange histograms: secondary stars at 1 Myr. Each histogram is normalized to unity.}
    \label{fig:hist}
\end{figure*}


Figure \ref{fig:hist} shows that NSs notably change their orbit in set~B, with respect to set~A and C. We detail the case of eccentricity: in set~A and C only $\sim{}16\%$ of NSs have eccentricities $e>0.5$, while in set B the fraction of NS with $e>0.5$ is as high as $\sim{}55\%$. On the other hand, $\sim{}25\%$ of secondary stars have eccentricities $e>0.5$ in set~A and C, while their fraction drops to $\sim{}5\%$ in set B. 
All the other orbital parameters show a similar behaviour, and show that  SN kicks are distributed between the two binary components, with weights that anti-correlate with the mass of the bodies.

{ Figure \ref{fig:erp} shows stellar pericentre distance versus 
eccentricity } for the three sets of runs, distinguishing between compact remnants and secondary stars. {  This Figure highlights that heavy objects are spatially more confined. }
 The bulk of stars and remnants still show orbital parameters that resemble the ones of the CW disc; however, several stars are scattered to highly eccentric orbits.

\begin{figure*}

\includegraphics[width=\textwidth]{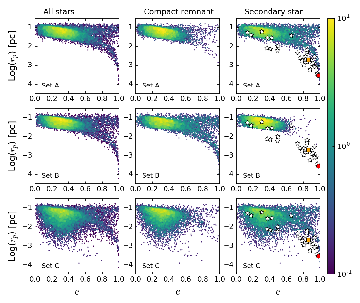} 
    \caption{{ Stellar pericentre distance ($r_{\rm p}$) versus eccentricity ($e$)} %
%
 of stars and compact remnants in set A (top), B (centre) and C (bottom), at the end of the simulations. The left-hand column shows all stars in each simulation set, the central panel shows only compact remnants, and the right-hand panel shows the secondary stars. For comparison, we plot the values of $e$ and $r_{\rm p}$  for G1 (orange square), G2 (red circle), and for the S-cluster members (white stars) over the scatter plot of secondary stars. The colour bar represents the density of points. }
    \label{fig:erp}
\end{figure*}

\begin{figure}
\includegraphics[trim={0cm 2cm 8cm 9.5cm},width=\columnwidth]{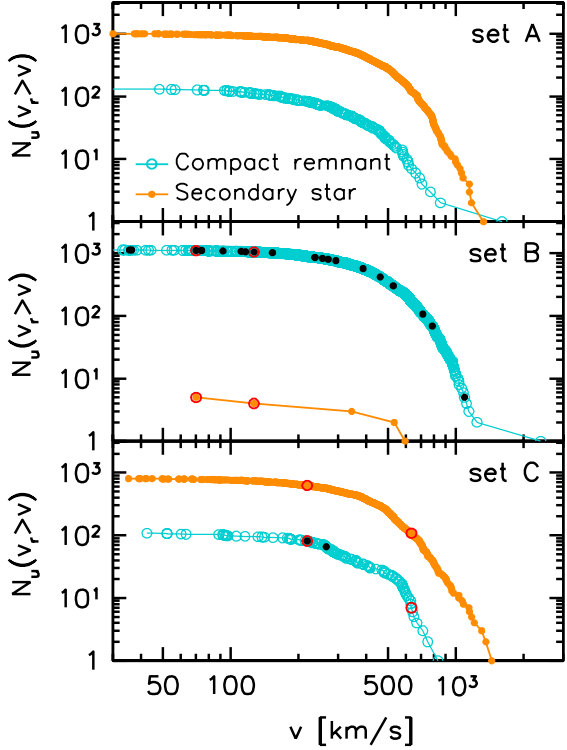} 
    \caption{Cumulative distribution of objects that escape (i.e. become unbound from) the SMBH, $N_u$, with radial velocity component $v_r$ greater than a threshold velocity $v$, as a function of $v$, at the end of the simulations. Light blue open circles: SN compact remnants; orange filled circles: secondary binary members. Escaping BHs are marked with a black filled circle. Bound binary systems that escape the SMBH are marked with a red circle. From top to bottom: cumulative radial velocity distribution for sets A, B and C. }
    \label{fig:escapers}
\end{figure}


\begin{table}
  \centering
  \caption{Fraction of objects that become unbound from the SMBH.}
  \label{tab:unbound}
  \begin{tabular}{lrrr} 
    \hline
     Object type &  set A & set B & set C \\
     \hline
         all     &   6\%  & 6\%   &    5\%  \\
         BHs     &   0\%  & 0.4\% & 0.04\%  \\
         NSs     &   3\%  & 24\%  &    2\%  \\
         II star &  10\%  & 0.05\%&    8\%  \\     
    \hline
\end{tabular}
\begin{flushleft}
{\footnotesize
Column 1: `Object type' indicates if a given object is a BH, a NS, a secondary (II) star or all possible objects.
Columns 2, 3, and 4: `set A', `set B', and `set C' indicate the statistics for each set of runs. 
Per each set, we list the fraction of unbound bodies over their total number, the relative fraction of unbound BH, NSs and secondary stars.}
\end{flushleft}
\end{table}

Most of stars and compact remnants continue orbiting the SMBH by the end of the simulation, and only  $\sim{}5\%$ objects become unbound from the SMBH. In set~B, NSs absorb most of the SN kick and about one fourth of them escapes the SMBH potential, while secondary stars are strongly anchored to their initial orbit and all but five of them stay on a bound orbit by the end of the simulation. We find an opposite trend in sets A and C, where secondary binary members are generally lighter  than NSs: about  10\% companion stars end up unbound, while only a few per cent of NSs escape. Only few BHs become unbound from the SMBH: 0\%, 0.4\%, and 0.04\% in sets A, B and C, respectively. Table~\ref{tab:unbound} is a summary of these possible outcomes.

Figure~\ref{fig:escapers} shows the cumulative distribution of radial velocities of stars and compact remnants that escape the SMBH potential well. One to a few per cent of these objects attain escape velocities $>900$ km s$^{-1}$, thus they may potentially be detected as hyper-velocity stars escaping the galactic potential (a more accurate investigation was already done by \citealt{Zubovas2013}). In set A and C, hyper-velocity stars are secondary binary members (except for one NS in set A). In set B only five  secondary stars escape from the SMBH, while the NSs are the fastest escapers. In four cases, both binary members escape the SMBH potential and they stay bound in a binary system.

\begin{table}
  \centering
  \caption{Fraction of stellar binaries that do not break 1~Myr after the SN kick.}
  \label{tab:bin}
  \begin{tabular}{lrrr} 
    \hline
     Object type &  set A & set B & set C \\
     \hline
           all   & 38\%   &  48\% &  39\%  \\
           BHs   & 69\%   &  80\% &  71\%  \\
           NSs   &  2\%   &   11\% &   2\%  \\     
    \hline
\end{tabular}
\begin{flushleft}
{\footnotesize
Column 1: `Object type' indicates if we consider all the binaries in the simulation, the binaries with a BH or the ones with a NS. Columns 2, 3, and 4: `set A', `set B', and `set C' indicate the statistics for each set of runs. 
Per each set, we list the fraction of objects that are still bound in a binary stellar system 1~Myr after the SN explosion; we list the fraction of survived binaries over the total initial number of binaries and the fraction of BHs or NSs in a bound binary normalized to the total of BHs or NSs in our simulations.}
\end{flushleft}
\end{table}

Stellar binaries break after the SN explosion in the { $50-60\%$} of runs. Table~\ref{tab:bin} lists the fraction of binaries that survive the SN kick. A higher fraction of binaries do not break in set~B, because secondary stars are more massive than compact remnants; however secondary stars will undergo SN explosion in a short timescale, thus even these binaries might be disrupted in few Myr.

\section{Discussion}\label{sec:disc}

\subsection{The population of dark remnants}

{ Our results show that in all simulation sets SN explosions are not effective in scattering massive objects on new orbits. If we assume continuous star formation in the inner pc, our simulations predict that a dark cluster of BHs may have {formed}  within the influence radius of the SMBH, because natal kicks do not modify the BH orbital properties  significantly.}


In set B, secondary stars (which are particularly massive) behave in a similar way as BHs, dominating the mass distribution within $\sim 0.02-0.5$ pc. However, almost all these high-mass stars will undergo a SN explosion within a few Myr and will get a further kick, changing their density distribution around the SMBH. The study of their subsequent evolution is beyond the aim of this paper; however we can speculate about their possible final distribution. It is reasonable to expect that a considerable fraction of the heavier BHs generated from secondary stars will keep moving in the same region, as SN kicks are expected to be weaker if the remnant is a BH. As a consequence, we expect the BH { population} of set B to acquire further members after the SN explosion of secondary stars. 
Thus, our simulations clearly show that { an excess} of dark remnants may hide inside the central parsecs of the Galaxy.

\begin{figure}
\includegraphics[trim={0cm 2cm 8cm 9.5cm},width=\columnwidth]{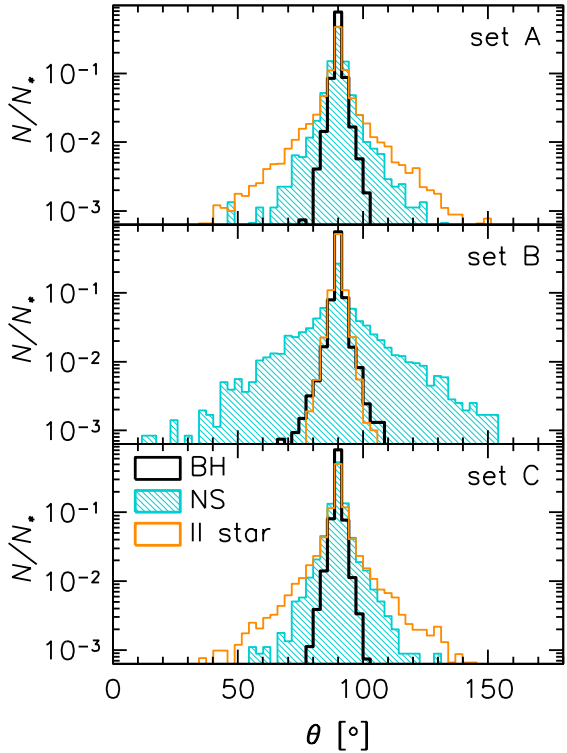} 
    \caption{Azimuthal angle ($\theta$) distribution of secondary stars and compact remnants still bound to the SMBH at the end of the integration. The initial disc is located at $\theta=90^{\rm o}$. Top panel: set~A; central panel: set~B; bottom panel: set~C. Black histograms: BHs at 1 Myr; light blue dashed histograms: NSs at 1 Myr; orange histograms: secondary stars at 1 Myr. Each histogram is normalized to unity.}
    \label{fig:theta}
\end{figure}


The angular distribution of objects resembles the initial one, i.e. all stars and remnants are preferentially found in the plane of the CW disc, even if light stars spread more their angular distribution, as shown in Figure \ref{fig:theta}; thus the final distribution of stars and compact remnants is flattened in our simulations. 
However, several star formation episodes might have occurred in disc-like structures (such as the CW disc) around the SMBH, with possibly different orientations. This might have contributed to build a { cluster} of dark remnants along the history of the GC. Several episodes of star formation within randomly oriented discs would result in a relatively isotropic distribution of remnants; in addition, secular processes as Kozai-Lidov oscillations \citep{Kozai1962,Lidov1962} or resonant relaxation \citep{Rauch1996} may have assisted the isotropisation of orbits.

The presence of a cusp within the inner parsec of our Galaxy can also give new hints on the origin of S-stars: \citet{Hills1988} proposed that both the S-star cluster and hyper-velocity stars within the galactic bulge may have originated by the tidal breakup of binaries passing close to the SMBH; \citet{Subr2016} investigated this possibility via $N$-body simulations, and they reproduce fairly well the orbital parameters of S-stars when they take in account the presence of a possible stellar or dark cusp.

\subsection{The NS distribution}

In our simulations, NSs are easily scattered far away from the SMBH, especially when they are bound to a massive companion. SN kicks are believed to have isotropic orientations; our simulations show that NSs have a greater probability to end up further away from the SMBH rather than closer in, as expected from geometrical arguments
\footnote{ If we take a sphere $S$ of radius $r$, centered on the SMBH and passing through the position $P$ of the exploding SN, and we consider the family of straight lines through $P$ (i.e. all possible directions of the SN kick), half of them is found to  avoid the sphere $S$; the other half will enter $S$, but it has to exit the sphere after covering a  distance in the range ($0,2r$).   
This means that if we consider all the
possible directions of the kick, in most cases the NS will end up farther away from the SMBH.}
. NS natal kicks may attain high velocities according to \citet{Hobbs2005}; in our simulations, we implicitly assumed that SN kicks have a constant distribution in momenta rather than in velocities, thus it is not surprising that NSs are heavily affected by their natal kick. We also expect that SN kicks will deeply affect NS orbits, even when they are not part of a binary system. Thus, we suggest that the dearth of NSs observed in the GC may be due to SN kicks.

Assuming a star formation rate in the GC of $\sim10^{-3}$ stars yr$^{-1}$ \citep{pfuhl2011} and introducing a correction factor to account for the fact that we did not simulate the low-mass end of the initial mass function (see Section~\ref{sec:methods}), we can compute the number of NSs that may { remain} in the GC within 10 Myr or 1 Gyr, i.e. the typical lifespan of respectively a canonical pulsar (CP) and a millisecond pulsar (MSP). Considering only sets A and C, we find that $\sim 500$ ($\sim 800$) NSs { would be located }  within the innermost 0.1 pc (1 pc) after 10 Myr, and $\sim 5\times 10^4$ ($\sim 8\times 10^4$) would be found in the same region after 1 Gyr.%
\footnote{ Note that the estimates we present are to be considered upper bounds on the number of NS that would { remain } inside 0.1 or 1 pc; as a matter of fact,  we neglected  NSs without a companion by assuming the binary fraction to be 1, but isolated NSs cannot share their natal kick with a companion and we expect them to spread farther away from the SMBH even more than NSs in binaries.} 

{
\citet{Rajwade2016} recently estimated the observable fraction of NSs in the inner pc of the Galaxy, and found that only the $\sim$2\% of CPs and $0.1\%$ of MSPs are nowadays detectable, even neglecting free-free absorption and multi-path scattering. They conclude that the maximum number of pulsars pointing toward us compatible with the null result of  present-day surveys is 50 CPs and 1430 MSPs,  suggesting that  we still do not have a definitive proof of missing pulsars. Following the calculations of \citet{Rajwade2016} and the results in the survey by \citet{Macquart2015}, if we assume that all our simulated NSs behave as CPs (MSPs), we would expect to observe $\sim$5 ($\sim 250$) of them at most; in addition, NSs can behave as MSP only in case they stay bound to the companion star or they dynamically acquire a massive companion; only 2\% of NSs in sets A and C do not break after the SN explosion, thus the number of candidate MSP in our sample could be reduced by a factor $\sim 10^{2}$. We will investigate  these aspects in more detail in a forthcoming paper.
}



\subsection{The S-cluster, G1 and G2}

\begin{figure*}

\includegraphics[width=.95\textwidth]{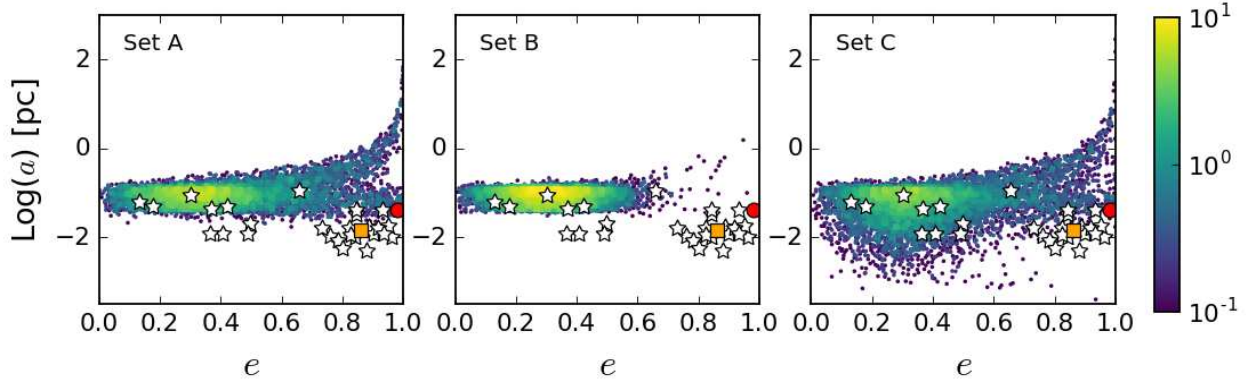} 
   \caption{Eccentricity ($e$) versus the semi-major axis
($s$) of secondary stars at the end of the simulations, in set A (left), B (centre) and C (right). For comparison, we plot the values of $e$ and $a$  for G1 (orange square), G2 (red circle), and for the S-cluster members (white stars). }
    \label{fig:ea}
\end{figure*}

\begin{figure*}

\includegraphics[width=.95\textwidth]{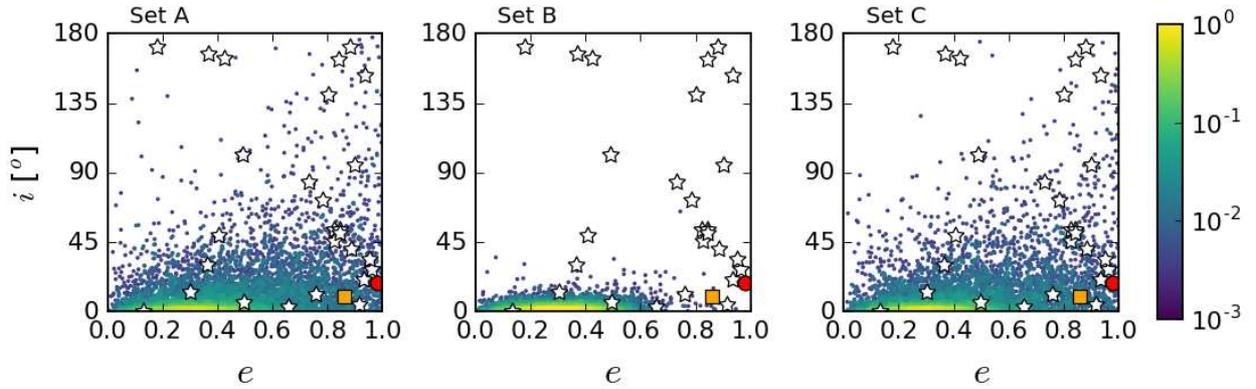} 

   \caption{Eccentricity ($e$) versus the orbital inclination
($i$) with respect of the CW disc of kicked secondary stars in set A (left), B (centre) and C (right), at the end of the simulations. For comparison, we plot the values of $e$ and $i$  for G1 (orange square), G2 (red circle), and for the S-cluster members (white stars). }
    \label{fig:ei}
\end{figure*}

The third column of Figure~\ref{fig:erp} and the three panels in Figure~\ref{fig:ea} and \ref{fig:ei} compare the orbital elements of secondary stars in our simulations  with those of the S-stars, G1 and G2. The scatter plots show the pericentre radii ($r_{\rm p}$), semi-major axes ($a$) and orbital inclinations ($i$) versus the orbital eccentricity ($e$), respectively. From these Figures it is apparent that the orbits of the S-stars with $e\lesssim{} 0.5$ may originate from the SN kick of a massive companion, especially in Set~C, where the inner edge of the disc is closer to the SMBH. 
However, most S-stars with eccentricities $\lesssim0.5$ are compatible with a CW-disc origin even without invoking a SN event. 

A second bunch of S-stars have eccentricity $e\gtrsim{}0.7$; the orbits of these S-stars (as well as those of G1 and G2) are marginally consistent with our distributions in sets A and C, but not in set B.  In particular, the 2.5\% (3\%) of secondary stars have $r_{\rm p}<10^{-2}$ pc and $e>0.7$ in our simulation set~A (set~C).  


 The orbital properties of a small fraction of the simulated stars in set A and C match those of G1 and especially G2. The fraction of secondary stars with eccentricity and semi-major axis compatible with those of G2 is 0.8\% and 1.0\% in set A and C, respectively. { The fraction of secondary stars with eccentricity and semi-major axis compatible with those of G1 is equal to 0.15\% for set C, while no stars match G1's orbit in set A}.
Low-mass stars experience a long pre-main sequence phase, and by the time they exit this phase a massive star born in the same star formation episode may experience the SN explosion; we propose that G1 and G2 are pre-main sequence stars that were orbiting a massive star in the CW-disc, and they were kicked on a highly eccentric orbit by the SN kick of the companion.

\subsection{X-ray binaries}

\emph{Chandra} X-ray observations of the GC and Galactic bulge show that  X-ray emitters are highly concentrated in the GC, in excess with respect to the expectations from stellar distribution models \citep{Muno2005,Hong2009}. This excess is mainly attributed to magnetic cataclysmic variables and primarily intermediate polars \citep{Hong2009}.  At least part of this  X-ray excess might originate from an accreting BH or NS binary.

In our simulations, a significant fraction ($40-50\%$) of stars is still part of a binary after the SN kick (Table~\ref{tab:bin}). To provide an order-of-magnitude estimate of how many of these simulated binaries might become X-ray emitters, we consider the minimum (min($R_\ast$)) and maximum (max($R_\ast$)) radius of each secondary star during its life, as obtained from the PARSEC stellar evolutionary tracks \citep{Bressan2012}. We then compare the minimum and maximum radii with the Roche lobe limit, obtaining a lower and upper limit for the number of binaries that may undergo Roche lobe overflow in our simulations. Our calculations show that $0-30\%$ of surviving binaries may undergo a phase of Roche-lobe overflow in sets A and C. The fraction of estimated Roche-lobe overflow systems is $0-76$\% in set B: if the secondary star is almost as massive as the primary, a non-negligible fraction of binaries in our simulation may become X-ray binaries (but we stress that our back-of-the-envelope calculation does not account for the lifetime of the companion star). Table~\ref{tab:xray} shows these results.  

{ \citet{Hong2009} found that the density of X-ray sources within a region of $40\times40$ pc$^2$ surrounding SgrA* is $\sim{}10^{-6}\msun^{-1}$, compatible with our generous upper limits. However, \citet{Hong2009} take into account all observed X-ray sources within an area much larger than the central pc, thus  a direct comparison with our results cannot be done. Moreover, to compare our results with observations we would need to account for binary evolution processes (such as Roche lobe overflow and wind accretion), which are not included in our simulations. A more accurate study of this aspect will be the subject of a forthcoming paper.}

\begin{table}
  \centering
  \caption{Fraction of stellar binaries that may turn on as X-ray emitters.}
  \label{tab:xray}
  \begin{tabular}{lrrr} 
    \hline
       & set A & set B & set C \\
     \hline
        min($R_\ast$)& 0 & 0.1\% &0.01\%\\
        max($R_\ast$)& 27\%   &  76\% &  29\%  \\
    \hline
\end{tabular}
\begin{flushleft}
{\footnotesize
Columns 1, 2, and 3: `set A', `set B', and `set C' indicate the statistics for each set of runs. 
Per each set, we list the fraction of stellar binaries surviving the SN kick that may turn on as X-ray sources, normalized to the total number of surviving binaries; the first line indicates a lower limit, the second line an upper limit in the possible X-ray emitters; see text for details.}
\end{flushleft}
\end{table}


\subsection{Caveats}

In our simulations, we did not take in account any secular processes. Mass precession may occur due to the spherical distribution of late-type stars in the GC; however this process would only cause pericentre advance and would let our results unaffected \citep{Subr2012}. Kozai-Lidov resonances \citep{Kozai1962,Lidov1962}  may have a role especially for stars with large semi-major axes; however Kozai cycles are expected to occur for stars misaligned to the CW disc; in addition, the spherical distribution of old stars would damp the cycles even for stars with inclined orbits. The net effect we may expect is a change in the longitude of the ascending node and argument of the pericentre, but again our results would probably be unchanged \citep{Mapelli2013,Trani2016b}. Moreover, two-body encounters can significantly affect the distribution of eccentricities and semi-major axes (e.g. \citealt{Trani2016b}).

A second possible issue of our simulations is the dependence of our results on the adopted kick distribution. The SN kick distribution is still highly debated (e.g. \citealt{Beniamini2016}, and references therein). We aim at investigating this issue in a future paper. However, we notice that our main result (i.e. the formation of a dark cusp) remains mostly unaffected by the prescription we adopt for the kick. In fact, recent studies \citep{Fryer2012,Mapelli2013b} suggest that BH kicks might be close to zero for all direct-collapse BHs, while our model attributes quite large kicks to BHs, even if they come from direct collapse. In this respect, our results represent a lower-limit for the building-up of a dark cusp.

\section{Summary}\label{sec:sum}

In this paper, we investigate the possibility that SN kicks occurring in binary systems affect the orbits of stars and stellar remnants in the GC, by means of regularized three-body simulations. We performed three main sets of simulations, changing the mass distribution of the secondary component of the binary and the inner edge of the initial disc. 

Our results show that BH remnants do not change their orbits significantly. Thus, several episodes of star formations in the GC may have built-up a cusp of BHs within the radius of influence of the SMBH. In contrast, NSs are generally scattered away from SgrA$^\ast$ and they may end up sweeping highly eccentric orbits; thus  the \emph{missing pulsars problem} may be a result of SN natal kicks, especially if a great number of NS progenitors are initially bound to a companion with a similar mass. Finally, we propose that G1 and G2 might have been low-mass stellar companions to massive CW-disc stars. When their massive companions underwent a SN explosion, G1 and G2 attained their highly eccentric orbits around the SMBH. This scenario deserves further investigation in a forthcoming paper.


\section*{Acknowledgements}
We warmly thank the anonymous referee for { his/her} useful comments and suggestions.  
We also thank Federico Abbate, James Petts, Alessia Gualandris,  Maureen van den Berg,  Melvyn B. Davies, Mark Gieles, John Magorrian, Alessandro Trani and Fabio Antonini for useful discussions and suggestions.  
We acknowledge the CINECA Awards N. HP10CP8A4R and HP10C8653N, 2016 for the availability of high performance computing resources and support.  We acknowledge financial support from the Istituto Nazionale di Astrofisica (INAF) through a Cycle 31st PhD grant, from the Italian Ministry of Education, University and Research (MIUR) through grant FIRB 2012 RBFR12PM1F, from INAF through grant PRIN-2014-14, from the MERAC Foundation and from the \textit{Fondazione Ing. Aldo Gini}.




\bibliography{biblio} 

\appendix
\section{Prescriptions for the natal kick on the binary members}\label{app}


{Here we describe the procedure we adopted to distribute the SN kick to the two members of the stellar binary system, according to the prescriptions 
detailed in appendix A1 by \citet{Hurley2002}.

If we consider a binary stellar system composed of two stars with initial masses $M_{1,i}>M_2$ and relative velocity vector $\bf r$, we can express their relative velocity as
\begin{equation}
{\bf v}=-v_0(\ihat \sin \beta +\jhat\cos \beta), 
\end{equation}
where $\beta$ is the angle between the position and velocity vectors; we use the notation $\ihat, \jhat,\khat $ to denote the unit vectors relative to a reference frame $X,Y,Z$ where the mass $M_2$ is in the origin and the specific  angular momentum $\vect{h}$ of the binary lies along the positive $Z$ axis. The system geometry is shown in Figure A1 of \citet{Hurley2002}. The angle $\beta$ can be expressed in terms of the orbital elements of the binary, i.e. the semimajor axis $a$, eccentricity $e$, eccentric anomaly E and the distance $r$ between the stars:
\begin{eqnarray}
\sin \beta & = & \left[ \frac{a^2 \left( 1 - e^2 \right)}{r \left( 2 a - r
\right)} \right]^{1/2}\\
\cos \beta & = & - \frac{e \sin {\rm E}}
{\left( 1 - e^2 {\cos}^2 {\rm E} \right)^{1/2}}.
\end{eqnarray} 
We chose the distance at which the SN explosion occurs by drawing a mean anomaly $\mathcal{M}$ (that is uniformly distributed in time) from a uniform distribution between 0 and 2$\pi$. For a given $\mathcal{M}$, we compute the associated eccentric anomaly by recursively solving
\begin{equation}
{\rm E}_{i+1}=\mathcal{M}+e\sin{\rm E}_i,
\end{equation}
where E$_i$ denotes the $i$-th iteration for the computation of E; the  tolerance for the convergence of E was set to $10^{-9}$. Then the distance $r$ between the stars can be computed as:
\begin{equation}
r=a(1-e\cos \rm E ).
\end{equation}
The orbital speed $v_0$ can as well be expressed as a function of the orbital elements:
\begin{equation}\label{eq:velapp}
v^2_0 = G M_{b,i} \left( \frac{2}{r} - \frac{1}{a} \right),
\end{equation}
where $M_{b,i}$ denotes the total mass of the binary before the SN explosion. 

We assume that the mass of the envelope, $\Delta M$, is instantaneously ejected by the primary star during its explosion, and its mass becomes $M_{1,f}=M_{1,i}-\Delta M$; thus the total mass of the binary after the SN is $M_{b,f}=M_{b,i}-\Delta M$. During the explosion, the distance between the stars does not vary, but the primary star experiences a velocity kick whose modulus $v_{k}$ is randomly drawn from a Maxwellian with one-dimensional variance equal to 265 km s$^{-1}$ \citep{Hobbs2005}; if the remnant is a BH (i.e. $M_{1,f}>3\msun$), the kick velocity is then normalized to the mass of the remnant, assuming linear momentum conservation. 
The kick velocity vector can be expressed as
\begin{equation}
\mathbf{v_k}=v_k(\ihat \cos \omega \cos \phi + \jhat \sin \omega\cos \phi + \khat \sin \phi);
\end{equation}
the direction of the kick is assumed to be isotropic in space, thus $\phi$ is distributed between $-\pi/2$ and $\pi/2$ as $P(\phi )\propto  \cos \phi $, while $\omega$ is uniformly distributed between 0 and 2$\pi$. After the SN explosion took place, the new velocity between the stars is:
\begin{eqnarray*}
\vect{v}_f & = & \vect{v} + \vect{v}_k \\ 
 & = & \left( v_k \cos \omega \cos \phi - v_0 \sin \beta \right) 
\ihat + \\ 
& & \left( v_k \sin \omega \cos \phi -  v_0 \cos \beta \right) 
\jhat + v_k \sin \phi \khat \, ,
\end{eqnarray*}
and its modulus is, according to equation \eqref{eq:velapp},
\begin{eqnarray}
v_f^2 & = & GM_{b,f}\left(\frac{2}{r}-\frac{1}{a_f}\right) \\
      & = & v_k^2 + v_0^2 - 2 v_0 v_k \left( \cos \omega 
\cos \phi \sin \beta + \right. \\ 
      & & \hspace*{3.0cm} \left. \sin \omega \cos \phi \cos \beta 
\right) \, . 
\nonumber 
\end{eqnarray}
Note that we use the subscript $f$ to denote the new binary orbital parameters after the SN kick. The previous equation can be solved to obtain the new semimajor axis $a_f$ of the binary system. The system  specific angular momentum after the SN is equal to
\begin{equation}
{\vect{h}_f} = \vect{r} \times \vect{v}_f,
\end{equation}
and its squared modulus is 
\begin{equation}
| \vect{r} \times \vect{v}_f |^2 =G M_{ b,f} a_f \left( 1 - e_f^2 \right),
\end{equation}
where 
\begin{equation}
| \vect{r} \times \vect{v}_f |^2 = r^2 \left[ v_k^2 {\sin}^2 \phi + \left( 
v_k \cos \omega \cos \phi - v_{\rm orb} \sin\beta \right)^2 \right] \, ;
\end{equation}
the previous equation can be solved to compute the new value for the eccentricity $e_f$. The new angular momentum vector is shifted with respect to the initial one by an angle $\eta$ such that
\begin{equation}
\cos \eta = \frac{v_0 \sin \beta - v_k \cos \omega \cos \phi}
{\left[ v_k^2 {\sin}^2 \phi + \left( v_k \cos \omega \cos \phi - v_0 
\sin \beta \right)^2 \right]^{1/2}} \, . 
\end{equation}

 Since an amount of mass $\Delta M$ is ejected from the primary star, the centre of mass of the system now exhibit a velocity shift equal to
\begin{equation}
\vect{v}_s = \frac{M_{1,f}}{M_{b,f}} \vect{v}_k - \frac{\Delta M M_2}
{{M}_{b,f} M_{b,i}} \vect{v} 
\end{equation} 
with respect to the initial reference frame.

}

\bsp	
\label{lastpage}
\end{document}